\documentclass[10pt,twocolumn,letterpaper]{article}

\usepackage[pagenumbers]{cvpr} 

\usepackage{graphicx}
\usepackage{amsmath}
\usepackage{amssymb}
\usepackage{booktabs}
\usepackage{caption}
\usepackage{subcaption}

\usepackage[pagebackref,breaklinks,colorlinks]{hyperref}

\usepackage[capitalize]{cleveref}
\crefname{section}{Sec.}{Secs.}
\Crefname{section}{Section}{Sections}
\Crefname{table}{Table}{Tables}
\crefname{table}{Tab.}{Tabs.}

\def\cvprPaperID{*****} 
\def\confName{CVPR}
\def\confYear{2023}

\begin{document}

\title{Thermal Analysis for NVIDIA GTX480 Fermi GPU Architecture}

\author{Savinay Nagendra\\
Department of Computer Science\\
The Pennsylvania State University\\
{\tt\small sxn265@psu.edu}
}
\maketitle

\begin{abstract}
   In this project, we design a four-layer (Silicon|TIM|Silicon|TIM), 3D floor plan for NVIDIA GTX480 Fermi GPU architecture and compare heat dissipation and power trends for matrix multiplication and Needleman-Wunsch kernels. First, cuda kernels for the two algorithms are written. These kernels are compiled and executed with the GPGPU Simulator to extract power logs for varying tensor sizes. These power logs are converted to ptrace files with an automation script written in Python. The 3D floor plan, along with the generated ptrace files are given to HotSpot, which generates thermal heat maps to show heat dissipation for various components of the Fermi architecture. These heat dissipation trends for both the kernels are observed for multiple tensor sizes to draw qualitative conclusions. The behavioral and execution patterns of both the kernels are also observed with these varying heat dissipation trends. With this project, we observe that increase in tensor size results in increase of heat dissipation in components of the Fermi Architecture. However, the temperature of the chip remains saturates after a particular tensor size, and remains constant thereafter. Heat dissipation is non-uniform with smaller tensor sizes, and becomes more uniform after a certain tensor size. This means, after a particular tensor size, more cores of the architecture get activated in the computations, thereby resulting in an almost constant temperature. We also observe that Needleman Wunsch uses more data movement between DRAM and caches, thereby showing higher heat dissipation patterns in DRAMs when compared to Matrix multiplication for the same tensor size. Our observations are in accordance with the theoretical concepts behind the working of the two algorithms, thereby making our results consistent. The link to the project files and scripts is \url{https://drive.google.com/file/d/1ukmSbWnk4ctFHF87x3lNsMU5qijbEAPu/view?usp=share_link}
\end{abstract}

\section{Design}
In this section, we will discuss the system design for thermal analysis. 
\subsection{Tools or Software packages Used}
In this section, we will discuss the tools used in this project.
\subsubsection{GPGPU-Sim} GPGPU-Sim \cite{fung2007dynamic} is a micro-architecture simulator that provides a detailed simulation model of a contemporary GPU. GPGPU-Sim simulates a kernel by first transferring data to GPU memory. Then, the GPU kernels run on the GPGPU-Sim which reports statistics for the kernels. Finally, the data is transferred back to CPU memory. This simulator provides mechanisms for more efficient SIMD branch execution on GPUs \cite{nagendra2017comparison, nagendra2022constructing, funk2018learning,liu2021new,pei2021utilizing,nagendra2020cloud,nagendra2020efficient,nagendra2022threshnet,nagendra2024patchrefinenet,nagendra2023estimating,zhu2022rapid}. This simulator is capable of dynamically regrouping threads into new warps on the fly following the occurrence of diverging branch outcomes. Power efficiency has become a more crucial evaluation metric than peak performance for mainstream GPGPU computing. 
\begin{figure}[htpb]
    \centering
    \includegraphics[width=0.7\linewidth]{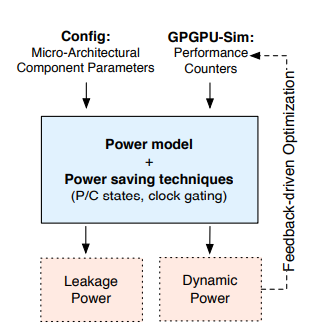}
    \caption{GPU Watch Power model integrated with GPGPU-Sim}
    \label{fig:gpgpu}
\end{figure}
GPU Watch \cite{leng2013gpuwattch} as shown in \cref{fig:gpgpu} is a GPGPU power model that is configurable, capable of cycle-level calculations, and carefully validated against real hardware measurements. It has been integrated with GPGPU-Sim. This power model estimates the power consumed by the GPU according to the timing behavior. This makes it ideal for evaluating fine-grained power management mechanisms. GPU watch uses a bottom-up method to build the initial model. The simulated power from this model is compared against the measured hardware power to identify modeling inaccuracies. These are resolved with a special suite of 80 microbenchmarks that are designed to create a system of linear equations that correspond to the total power consumption. In this project, we are using the integrated GPU Watch/GPGPU-Sim to generate the power report for our kernels. 

\subsubsection{HotSpot}
Hotspot \cite{hotspot} is a pre-RTL thermal simulator that is used in the early design process to analyse heating patterns of the designed architectures. It supports simulation of traditional 2D Integrated Circuits, 3D Integrated Circuits and Microfluiding Cooling. Hotspot takes as input a power trace file with power statistics of each component in the architecture and a 2D/3D floor plan of the architecture to generate heat dissipation maps of the architecture. The color variations indicate differential heating of different components, thereby allowing researchers to improve their designs. 

\subsection{Cuda Kernels}
In this section, we will discuss the implementation of matrix multiplication and Needleman-Wunsch kernels in written in cuda. In cuda, there is the concept of grids and blocks. Blocks are 3D arrays of threads and Grids are 3D arrays of blocks. Each thread knows the $x$ and $y$ coordinates of the block it is in the grid, and the coordinates of where the thread itself is in the block. These positions are used to calculate a unique thread ID for every thread. The total number of threads in a single block cannot exceed 1024.  
\subsubsection{Linearized Matrix Multiplication Kernel}
Assume we have two matrices $A$ and $B$, where $A$ is a $n \times m$ matrix and $B$ is a $m \times w$ matrix. Then, the result $M = A \times B$ is a $n \times w$ matrix. The first element of $M$ is the sum of all the element-wise multiplications of the numbers in the first row of $A$ with the numbers in first column of $B$. We have to compute every element in $M$, where each of them is \textbf{independent} from the others. This means, matrix multiplication can be effectively parallelized. We realized with experiments that regular matrix multiplication with three for loops is not an effective way of implementing the kernel. The simulation time this was taking was significantly high, not allowing us to use tensor sizes beyond $80 \times 80$. \begin{figure}[htpb]
    \centering
    \includegraphics[width=\linewidth]{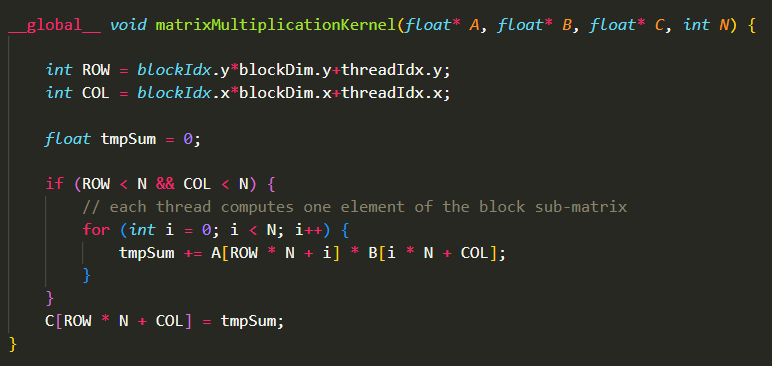}
    \caption{Linearized Matrix Multiplication Kernel \cite{matmul1}}
    \label{fig:matmul1}
\end{figure}So, we decided to make use of 1D arrays for our matrices. The insight is the way Cuda is developed, where the number of columns must be known before compiling the program. Linearizing the matrices allows us to know this, and dynamically calculate the rows and columns for indexing. Linearization was done by stacking each row length-wise from first to last. The result for an index is calculated by $C[row * N + col] = A[row * n + i] * B[i * N + col]$, where $i$ is the index in the linearized array (thread number) as shown in \cref{fig:matmul1} This cuda kernel was referred from \cite{matmul1}. The distribution of work between threads across blocks is make such that there is one thread  responsible for one $(i,j)$ index. So, the total number of threads is equal to the number of elements in the matrix. 

\subsubsection{Linearized Matrix Multiplication Kernel optimized by using Scratchpad memory (Cache Tiling)}
\begin{figure}[htpb]
    \centering
    \includegraphics[width=\linewidth]{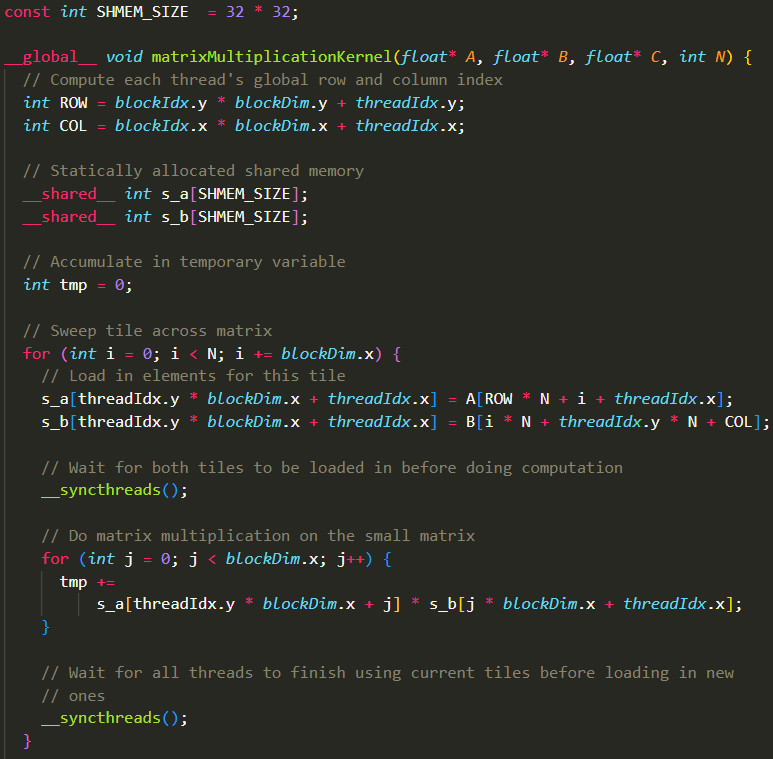}
    \caption{Linearized Matrix Multiplication Kernel optimized with Scratchpad memory.}
    \label{fig:matmul2}
\end{figure}
Without optimization, every single thread is marched along main memory, computing the product of two elements 
and then accumulating into the result. With Shared memory (scratchpad memory) optimization, we are going to do the same operation, but in chunks of thread block dimension ($32\times 32)$ elements. Each thread block will compute chunks of 32 elements of matrix multiplication at a time. This way, we know what exactly the memory we are going to be accessing next. We are exploiting this fact to load the next tile into the private cache of the Streaming Multiprocessor (SM). So, we will load in 32 chunks into the shared memory as shown in \cref{fig:matmul2} When all the threads are accessing those chunks, instead of going into the main memory and hoping to get lucky in the cache hierarchy, we can guarantee that the memory access will be in the private L1 level, making it very fast. This way, we get the performance speedup, DRAM accesses can be reduced. With this optimization, it can be observed that this kernel executes in significantly lesser amount of time when compared to the previous one, along with significantly lesser DRAM access. 

\subsubsection{Needleman-Wunsch Kernel}
\begin{figure}[htpb]
    \centering
    \includegraphics[width=\linewidth]{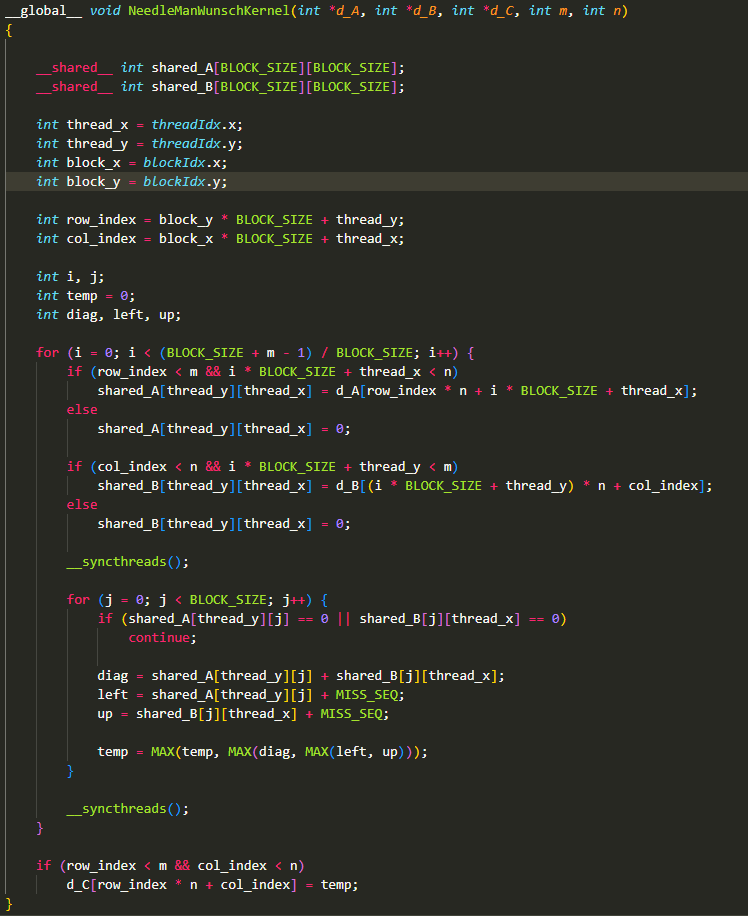}
    \caption{Needleman-Wunsch kernel optimized with Scratchpad memory.}
    \label{fig:nmw}
\end{figure}
\begin{figure}[htpb]
    \centering
    \includegraphics[width=\linewidth]{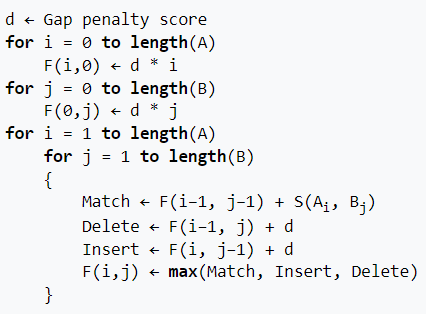}
    \caption{Needleman-Wunsch Psuedo Code}
    \label{fig:nwm_p}
\end{figure}

The Needleman-Wunsch algorithm is a 2D dynamic programming algorithm that is used in bioinformatics  to align protein or necleotide sequences. The algorithm divides the large problem into a series of smaller problems, and uses the solutions to the smaller problems to find the optimal solution to the larger problem. The cuda kernel shown in figure \cref{fig:nmw} uses the following psuedo-code for implementation as shown in \cref{fig:nwm_p}. The code for this kernel implementation is taken from \cite{nmw}.

\subsection{NVIDIA GTX480 Fermi}
In this section we will discuss the GTX480 fermi architecture \cite{glaskowsky2009nvidia} and floor plan design. 
\subsubsection{Fermi Architecture}
\begin{figure}[htpb]
    \centering
    \includegraphics[width=\linewidth]{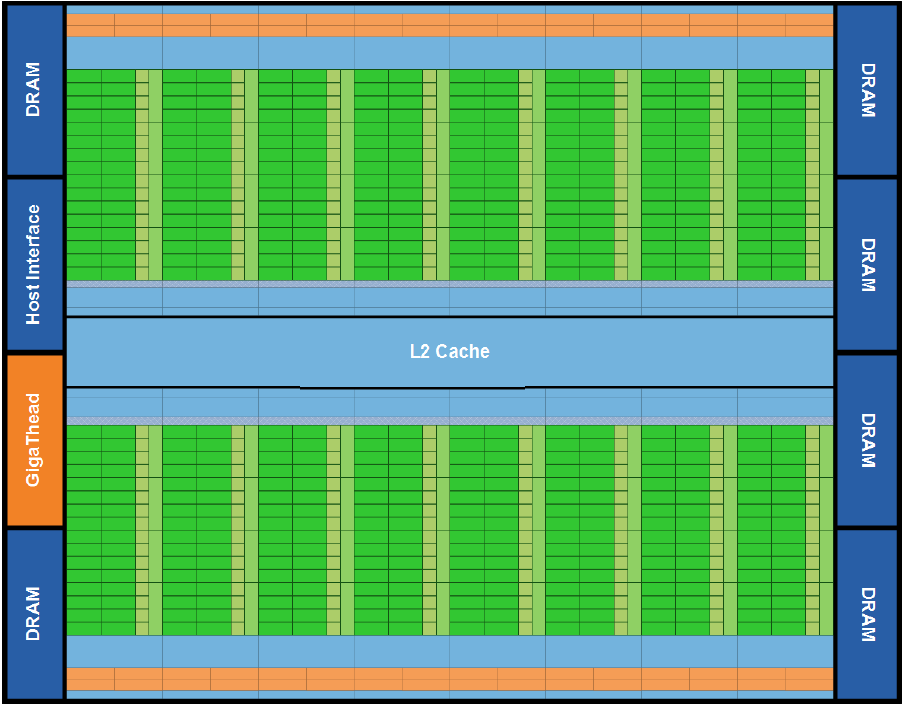}
    \caption{Fermi Architecture}
    \label{fig:fermi1}
\end{figure}

The Fermi architecture, as shown in \cref{fig:fermi1} has upto 512 CUDA cores organized in 16 Streaming Multiprocessors (SM) of 32 cores each, where each CUDA core executes a floating point or integer instruction per clock for a thread. The GPU has six 64-bit memory partitions, for a 384-bit memory interface, supporting upto a total of 6GB of GDDR5 DRAM memory. PCI-Express bus connects the GPU to the CPU. The GigaThread global scheduler is responsible for distributing thread blocks to SM thread schedulers. 
\begin{figure}[htpb]
    \centering
    \includegraphics[width=\linewidth]{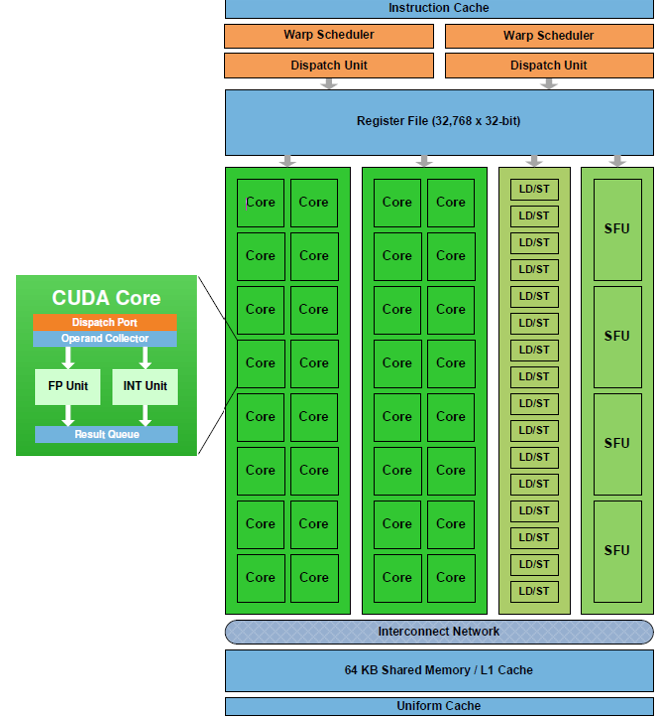}
    \caption{Fermi Streaming Multiprocessor}
    \label{fig:fermi2}
\end{figure}
The SMs are the execution units of the GPU. Each SM has 32 CUDA cores. Each CUDA processor has a fully pipelined ALU and FPU. Each SM has 16 load/store units, allowing source and destination addresses to be calculated for 16 threads per clock. Each SM has its own 64KB L1 cache, an uniform cache, an Instruction cache, a Scheduler-dispatcher unit and a $32768 \times 32-bit$ register file as shown in \cref{fig:fermi2}. 

\subsubsection{Fermi Floor Plan}
The floor plan was designed for TIM Layers 0 and 2 based on the diagram in \cref{fig:fermi1}. The chip dimensions were taken to be $23mm \times 23mm$. The dimensions for each component was proportionally calculated and put in two $.flp$ files, which were fed to HotSpot. Only the size DRAMs and the L2 cache were placed on Layer0. The rest of the components were made into a void placeholder. The 16 SMs and their respective components were place on Layer2. Only the components for which the power values are generated  by GPGPU-Sim were considered while designing the floor plan.

\section{Methodology}
\begin{figure}[htpb]
    \centering
    \includegraphics[width=\linewidth]{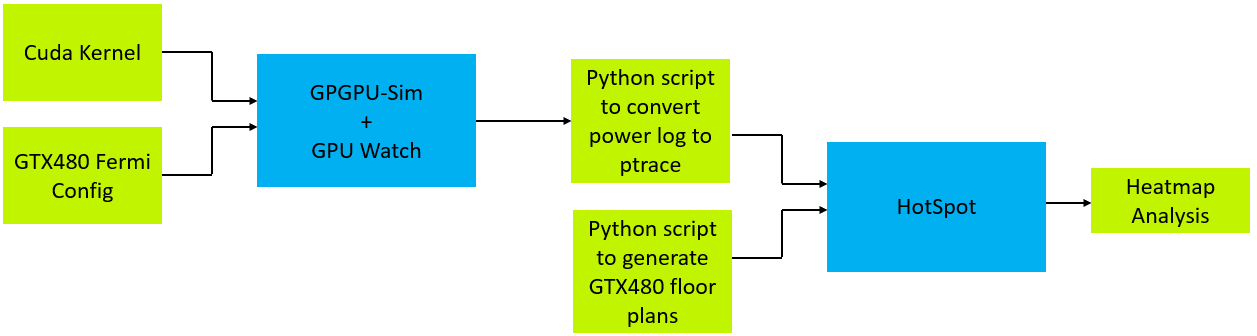}
    \caption{Block diagram showing the flow of this project. The green boxes indicate the work I have done.}
    \label{fig:blk}
\end{figure}

In this section, we will discuss the flow of the project. \cref{fig:blk} shows the flow of this project. First CUDA kernels as mentioned in the previous section are written.  
A kernel along with the GTX480 config file are given as inputs to the GPSPU-Sim + GPUWatch interface to generate power report. This is repeated for all the kernels and varying tensor sizes. A python script parses the power report to extract power values for each component of the Fermi architecture. The power report has three values - minimum, average and maximum, Two other values are added by using interpolation, such that the average of the five values remains the same as the previous average. The output of the python script is a $.ptrace$ file that will be given as an input to HotSpot tool. I have also written a python script to generate Layer0 and Layer2 floor plans as mentioned in the previous section. This script uses the proportionality of component dimensions in \cref{fig:fermi1} and generates two $.flp$ files. The $.ptrace$ file along with the two $.flp$ files are given as inputs to the HotSpot code. The $run.sh$ file is modified accordingly. Running this file will generate heat maps for the four layers. We only use/need the heat maps for Layer2 in all our results. We observe that Linearized Matrix Multiplication kernel does not provide significantly different results in terms of DRAM heat dissipation for increasing tensor sizes. Hence, we also decrease the L2 cache size in the config file (along with the corresponding L2 cache area in the floor plan) to observe the DRAM's heating patterns. Since heating patterns are more expressed with reduced L2 cache size, the cache tiling results are also taken for reduced L2 mode, so that we can compare how DRAM accesses are reduced with CUDA kernel optimization. While running Needleman-Wunsch kernels, we maintain the original L2 cache size. Matrix multiplication and Needleman-Wunsch kernels are varied for tensor sizes of 100, 250, 400 and 800. 

\section{Evaluation/Results}
In this section, we will observe and infer from the heat map dissipation patterns for the varying tensor sizes for the three matrix multiplication kernels.
\subsection{Linearized Matrix Multiplication Kernel}
\begin{figure}[htpb]
    \centering
    \includegraphics[width=\linewidth]{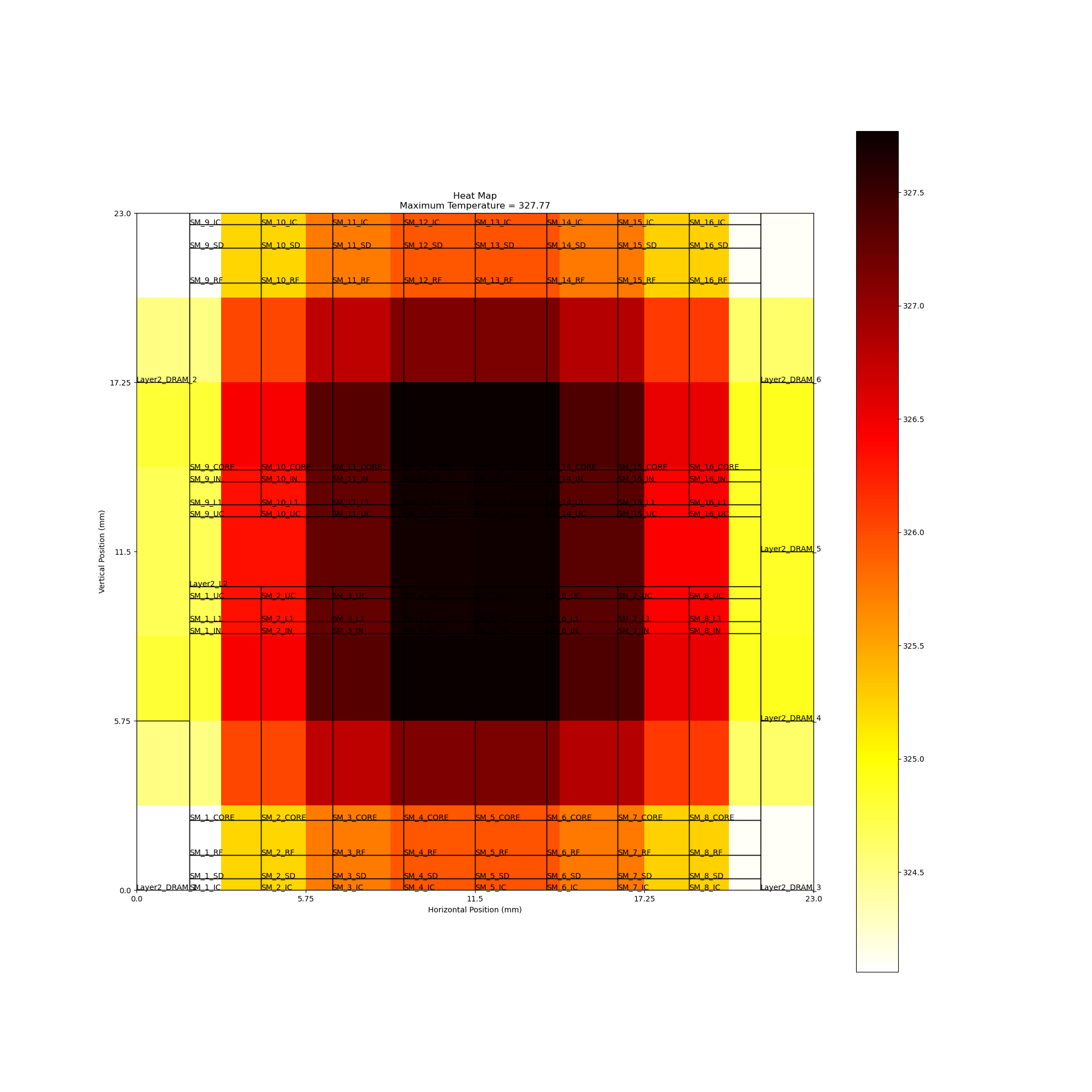}
    \caption{Linearized Multiplication Kernel: 100x100}
    \label{fig:11}
\end{figure}

\begin{figure}[htpb]
    \centering
    \includegraphics[width=\linewidth]{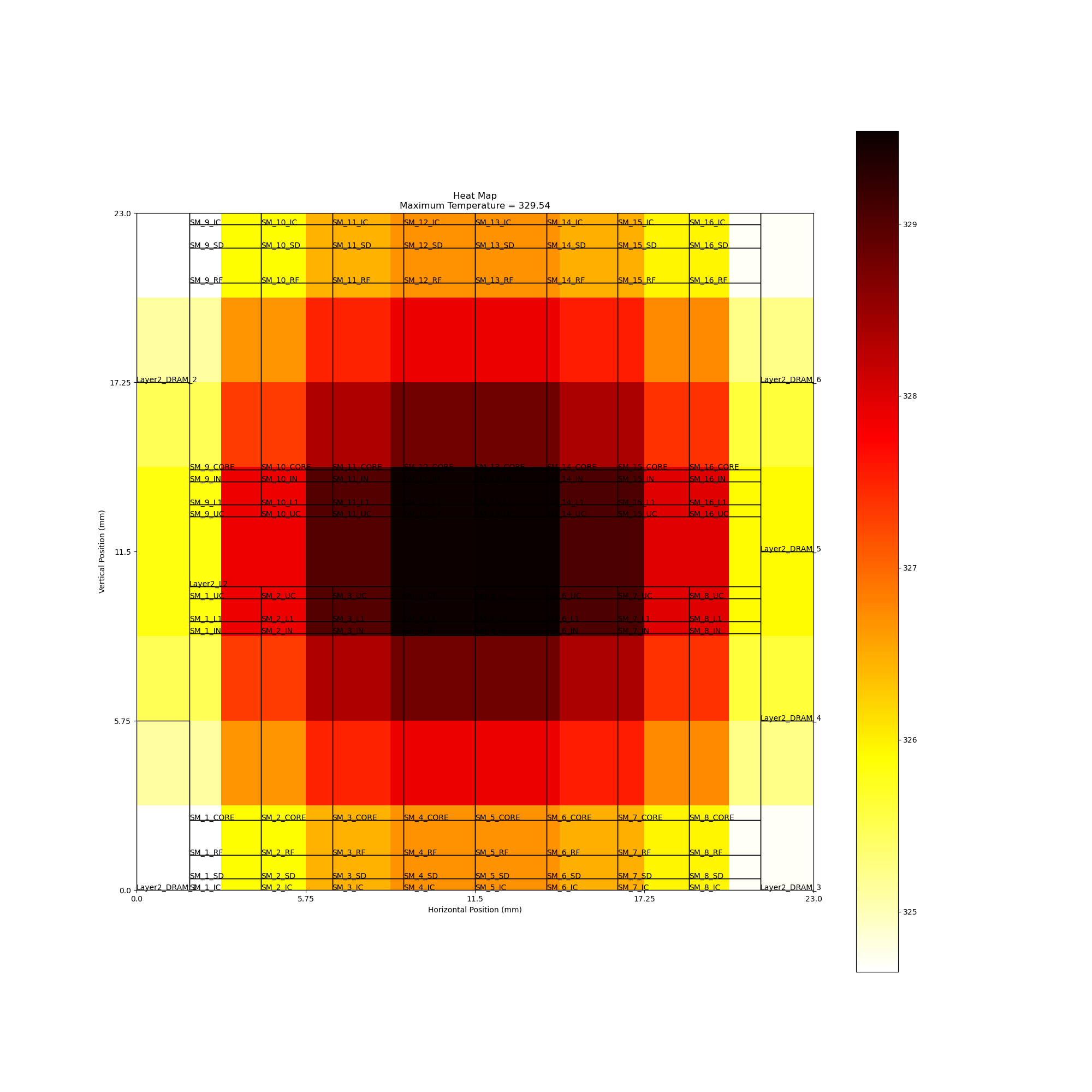}
    \caption{Linearized Multiplication Kernel: 250x250}
    \label{fig:12}
\end{figure}

\begin{figure}[htpb]
    \centering
    \includegraphics[width=\linewidth]{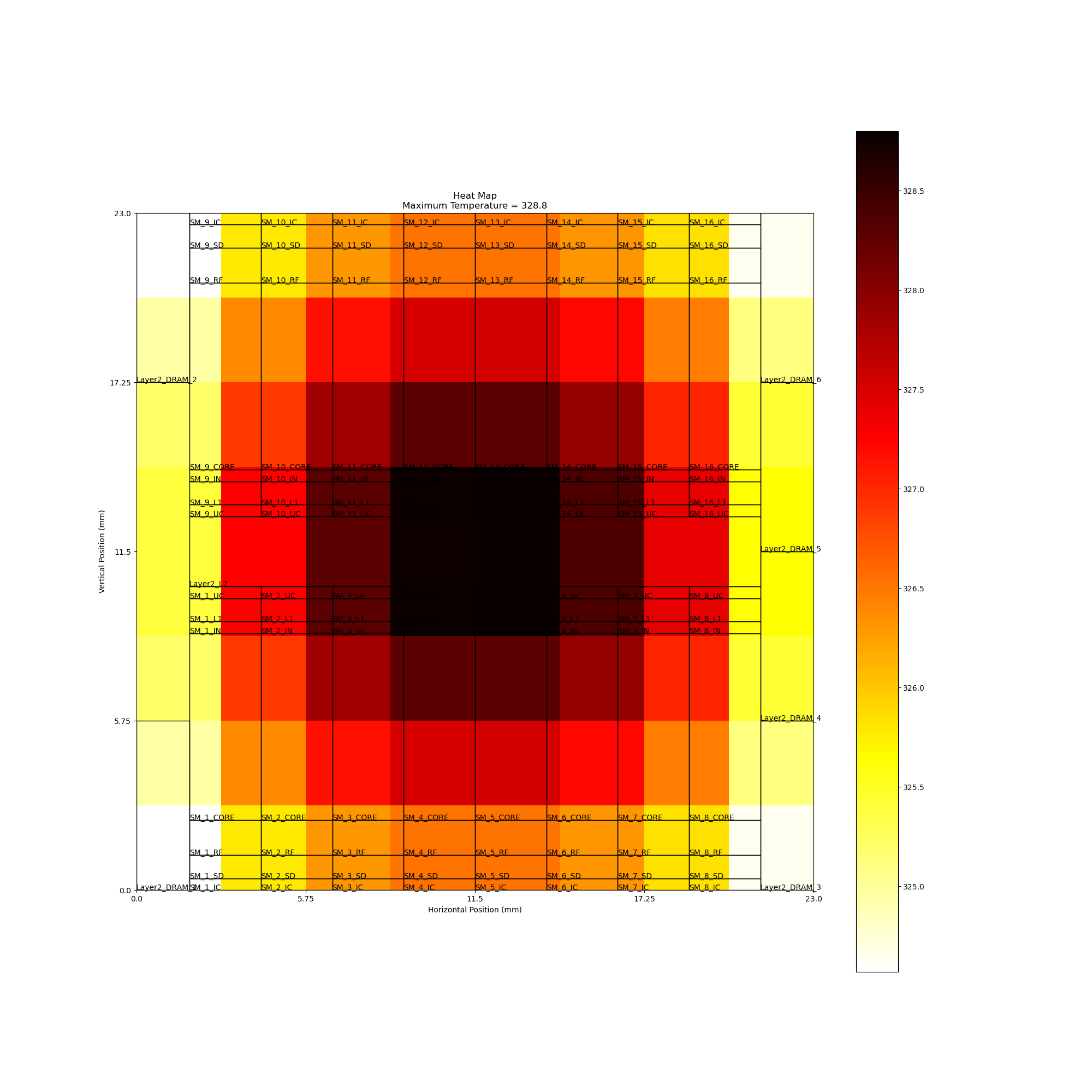}
    \caption{Linearized Multiplication Kernel: 400x400}
    \label{fig:13}
\end{figure}

\begin{figure}[htpb]
    \centering
    \includegraphics[width=\linewidth]{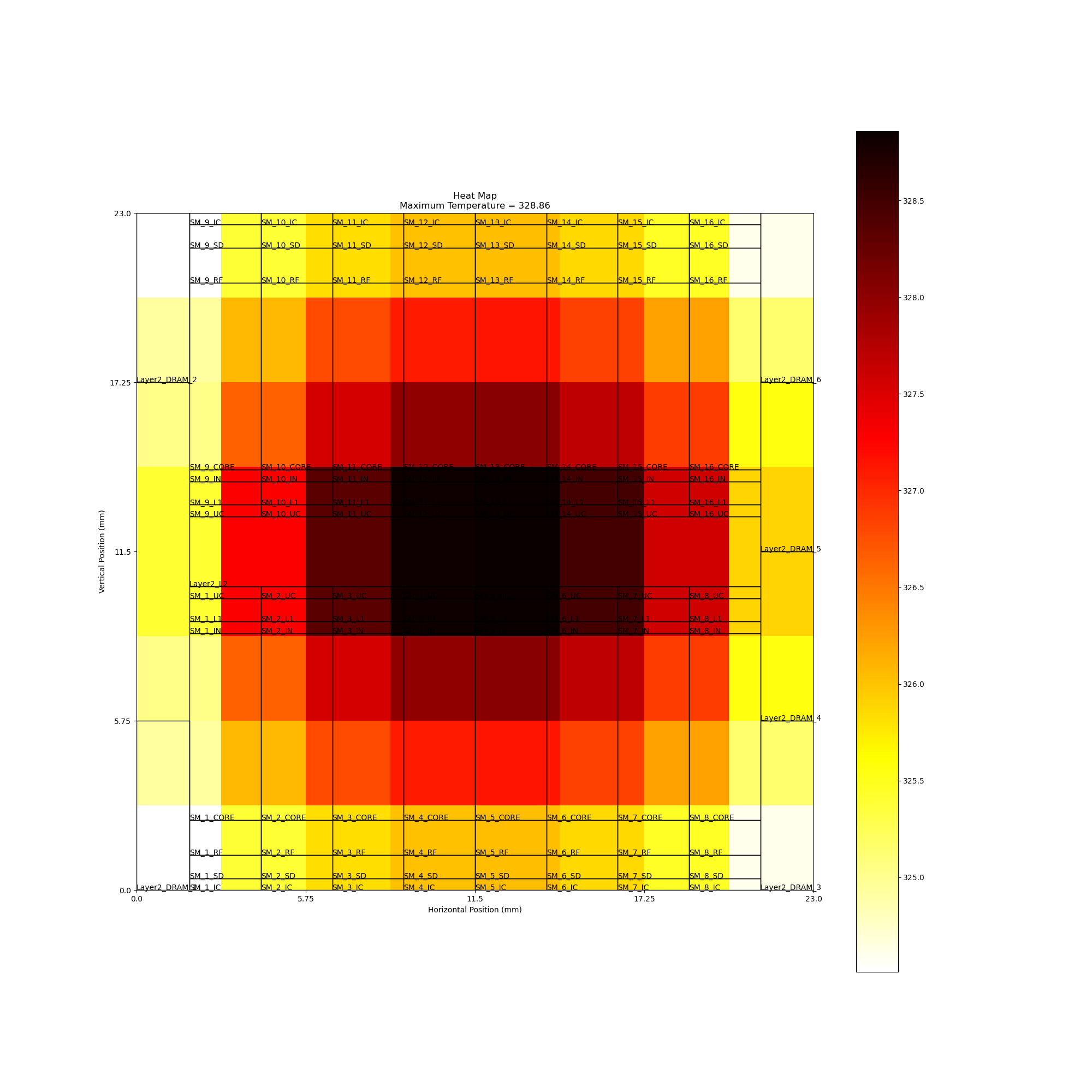}
    \caption{Linearized Multiplication Kernel: 800x800}
    \label{fig:14}
\end{figure}

In this section, we will look at the linearized matrix multiplication kernel's heat dissipation patterns. It can be observed that from figures \ref{fig:11} to \ref{fig:14} that with increase in tensor size from 100 to 800, the heat dissipated becomes more and more uniform, which leads to maintaining a constant temperature or even reduced temperature for the chip after $400 \times 400$. With an increase from $400\times 400$ to $800 \times 800$, it can be observed that the DRAM closest to L2 cache dissipates more heat. This is consistent with the theoretical observation where $800\times 800 \times 4 = 2500 KB$ data cannot fit on a $768KB$ L2 cache. Only this DRAM is used more so that the data movement between the DRAM and cache can be minimized. With increase in tensor size, more cores get activated, thereby making the heat dissipation uniform. 

\subsection{Linearized Matrix Multiplication Kernel with reduced L2 cache size}
\begin{figure}[htpb]
    \centering
    \includegraphics[width=\linewidth]{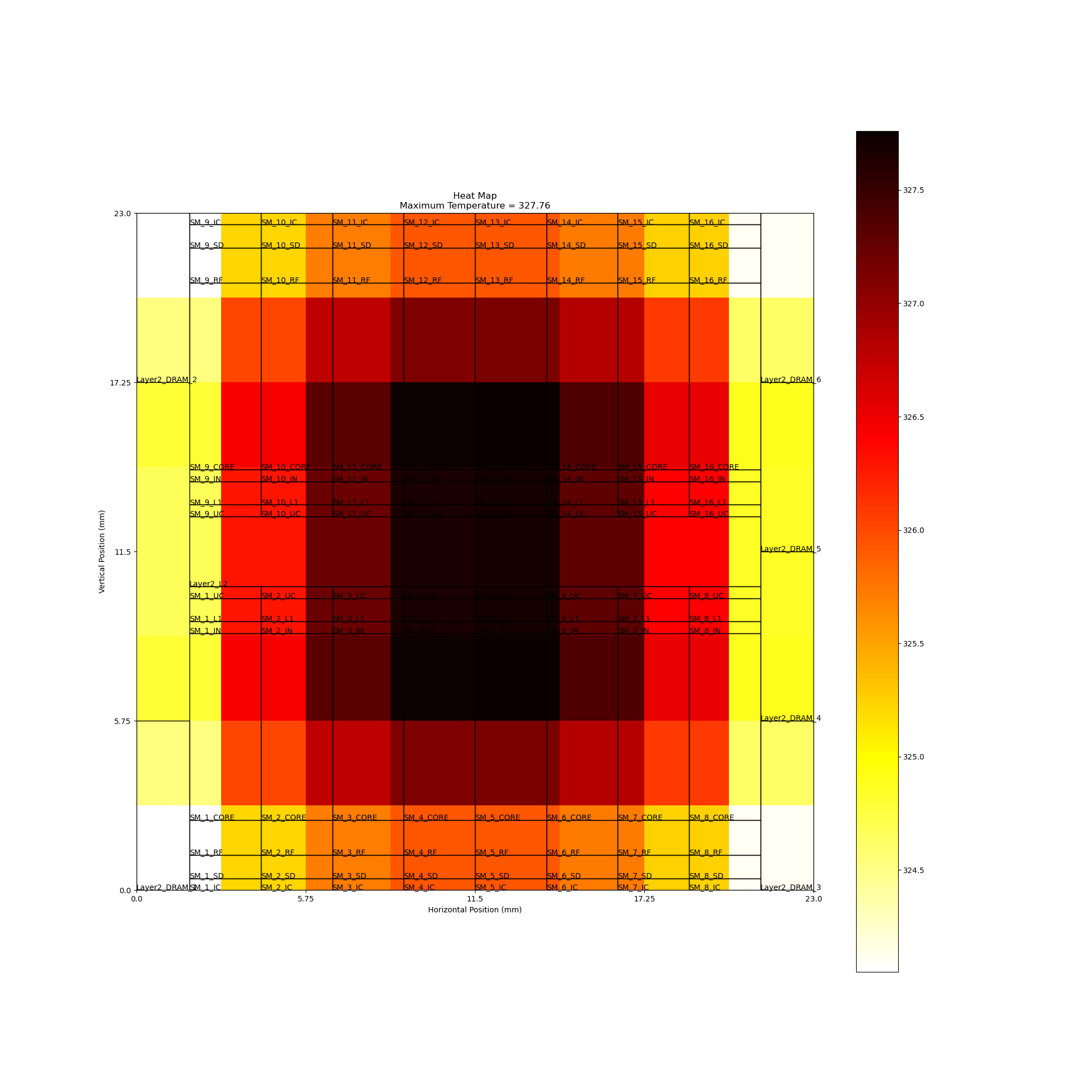}
    \caption{Linearized Multiplication Kernel with reduced L2: 100x100}
    \label{fig:21}
\end{figure}

\begin{figure}[htpb]
    \centering
    \includegraphics[width=\linewidth]{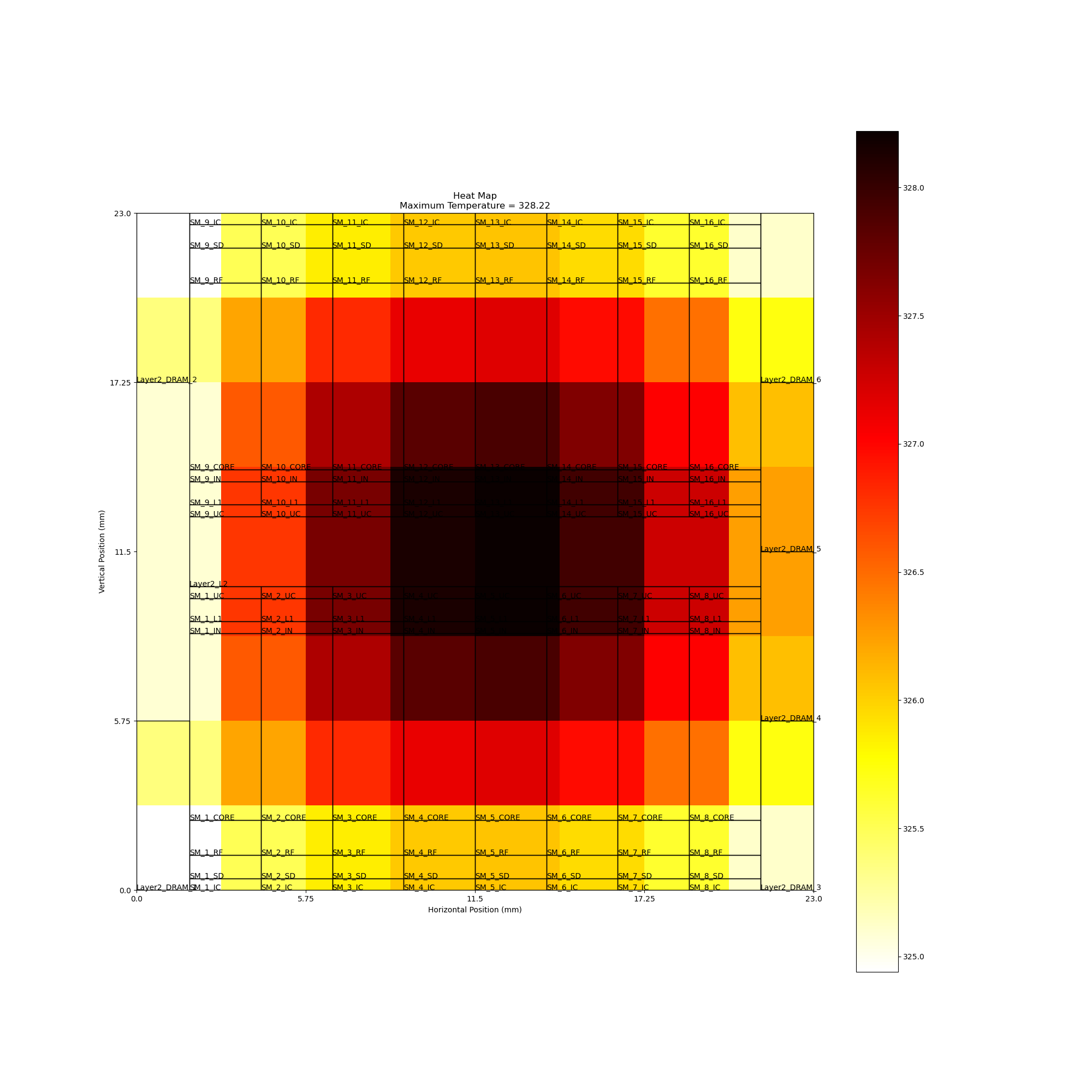}
    \caption{Linearized Multiplication Kernel with reduced L2: 250x250}
    \label{fig:22}
\end{figure}

\begin{figure}[htpb]
    \centering
    \includegraphics[width=\linewidth]{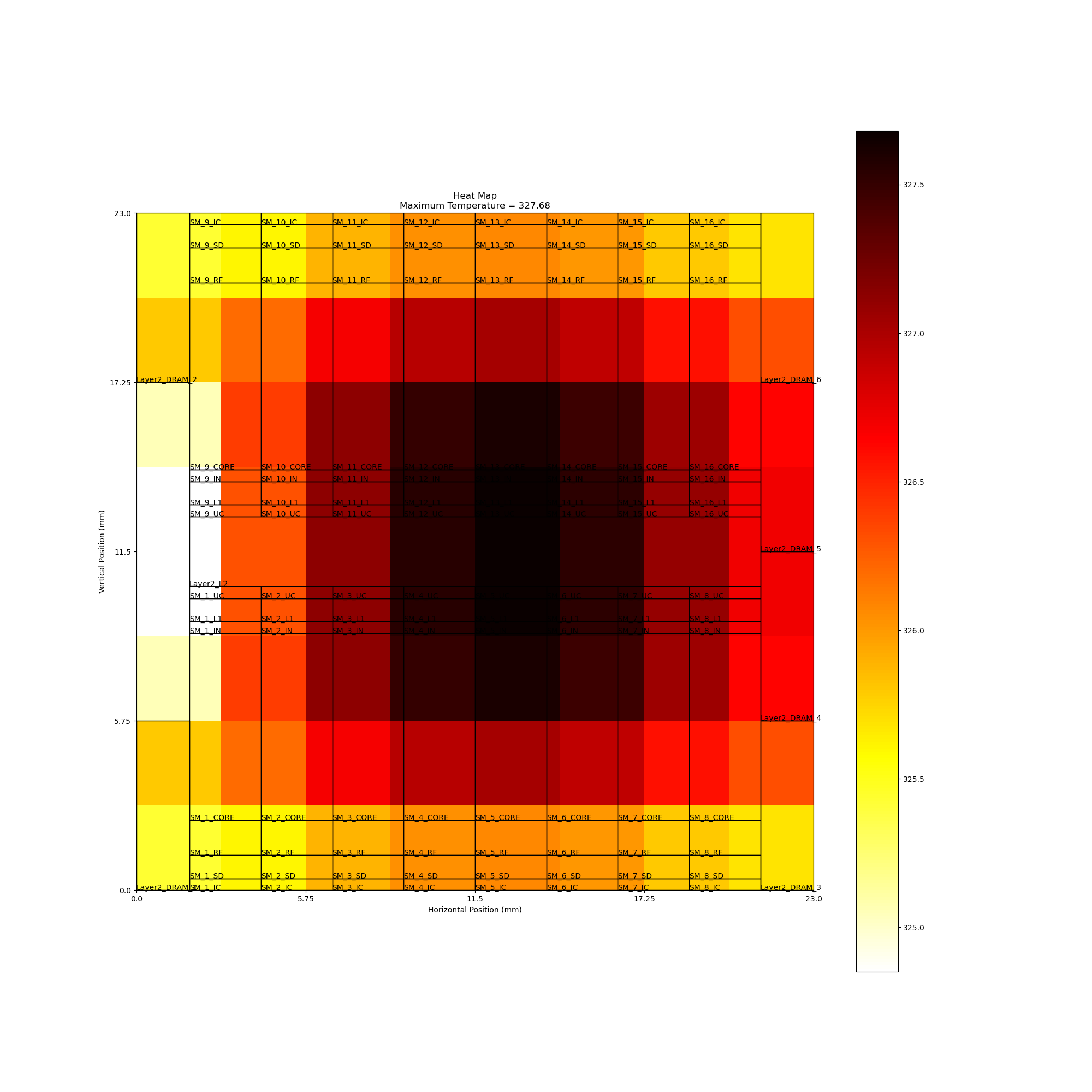}
    \caption{Linearized Multiplication Kernel with reduced L2: 400x400}
    \label{fig:23}
\end{figure}

\begin{figure}[htpb]
    \centering
    \includegraphics[width=\linewidth]{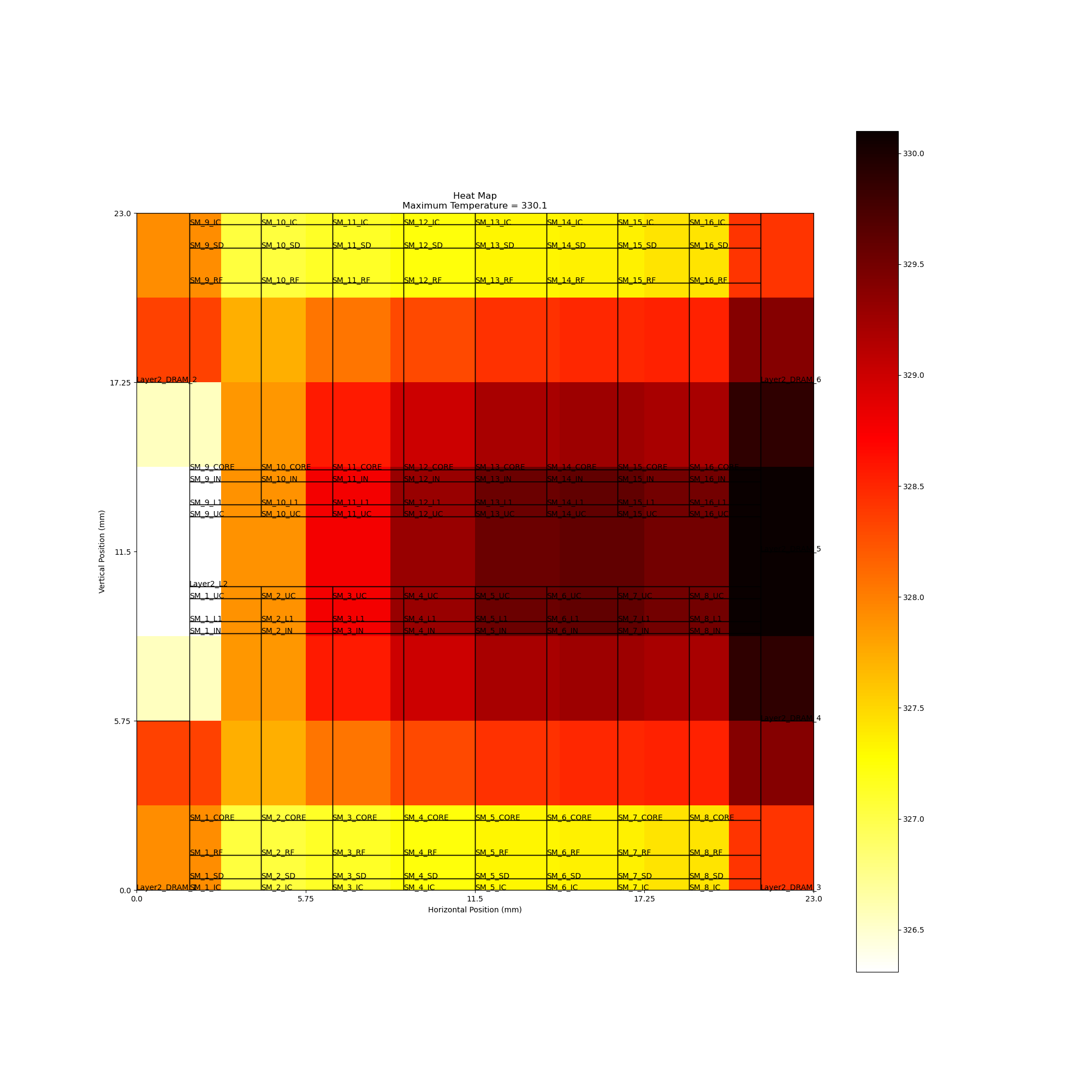}
    \caption{Linearized Multiplication Kernel with reduced L2: 800x800}
    \label{fig:24}
\end{figure}

Since execution times of the linearized kernel is high, I was only able to run matrix multiplication kernel for a maximum tensor size of $800 \times 800$. However, reducing L2 size provides the same effect as increasing tensor size . With this, I reduced the L2 size of the GPU to $\frac{1}{4}^{th}$ of its original value of $768KB$. The results are shown in figures \ref{fig:21} to \ref{fig:24}. It can be seen clearly that, with reduced L2 cache size, the GPU is now accessing DRAM significantly, leading to higher heat dissipation. Also, from \cref{fig:24}, it can be observed that all the three DRAMs on the right side are being used, leading to higher heat dissipation in each of them. However, the DRAM closest to L2 cache is heated up the most, as that is the one being used the most in order to minimize data movement. 

\subsection{CUDA optimized Linearized Matrix Multiplication Kernel with reduced L2 cache size}
\begin{figure}[htpb]
    \centering
    \includegraphics[width=\linewidth]{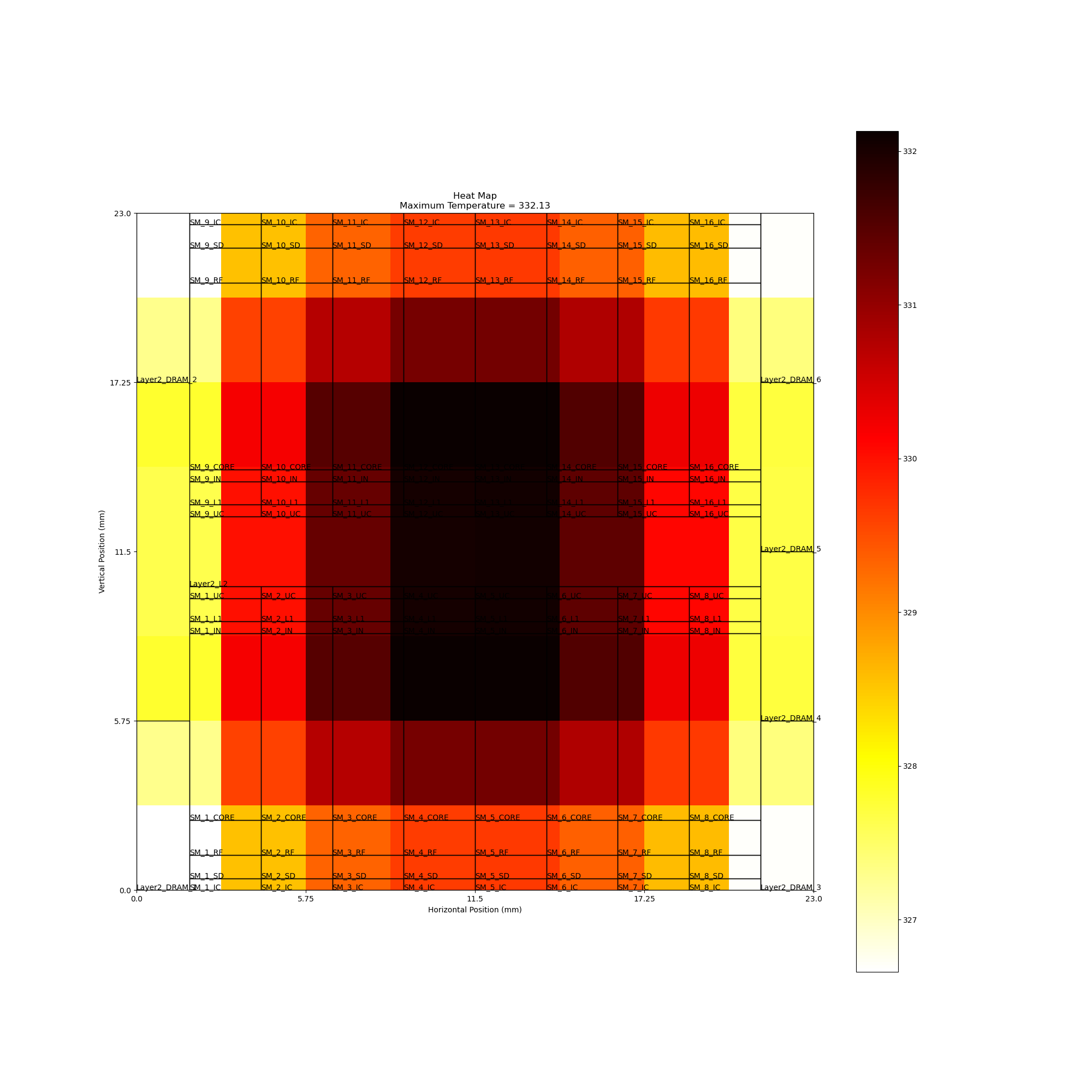}
    \caption{Cuda optimized Linearized Multiplication Kernel with reduced L2: 100x100}
    \label{fig:31}
\end{figure}

\begin{figure}[htpb]
    \centering
    \includegraphics[width=\linewidth]{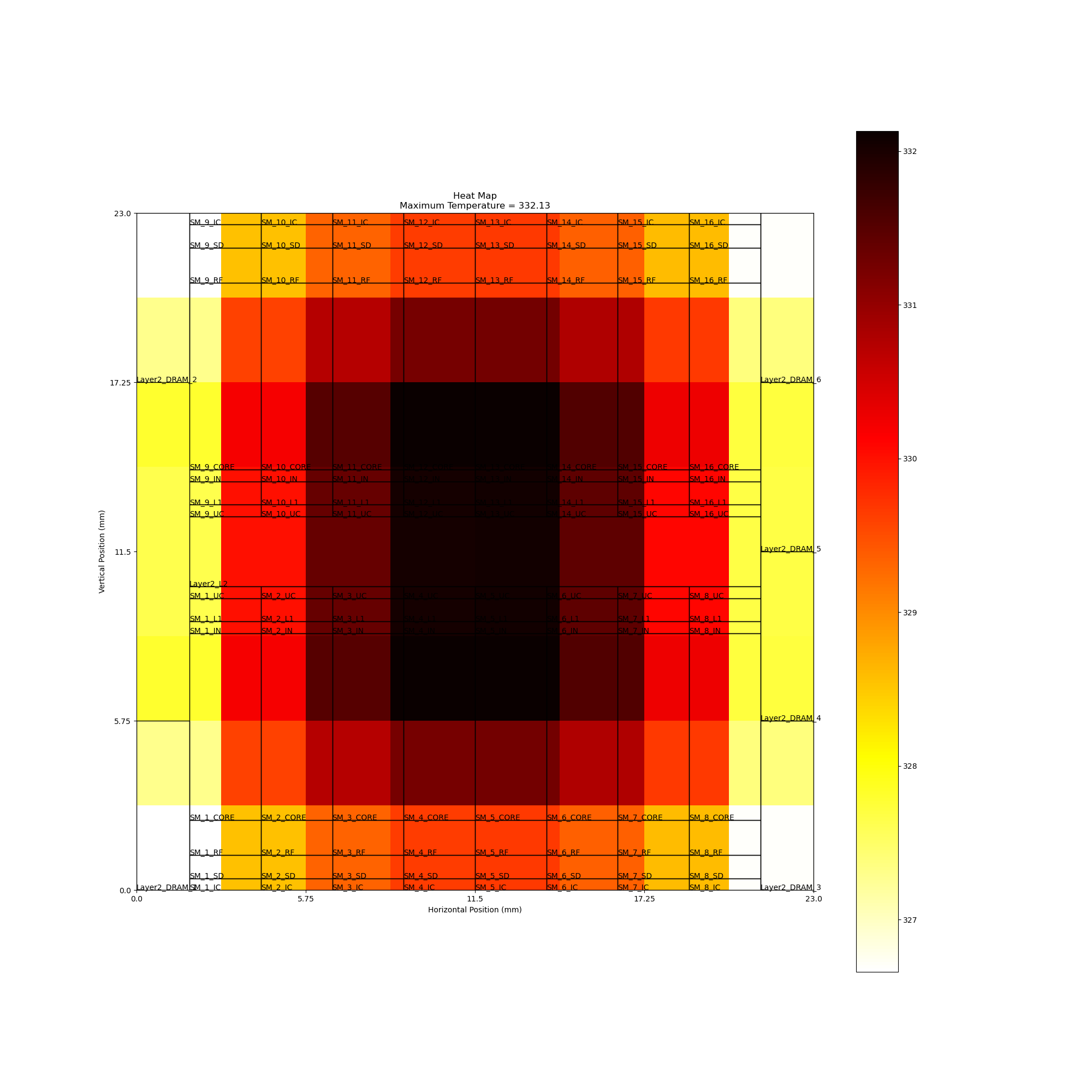}
    \caption{Cuda optimized Linearized Multiplication Kernel with reduced L2: 250x250}
    \label{fig:32}
\end{figure}

\begin{figure}[htpb]
    \centering
    \includegraphics[width=\linewidth]{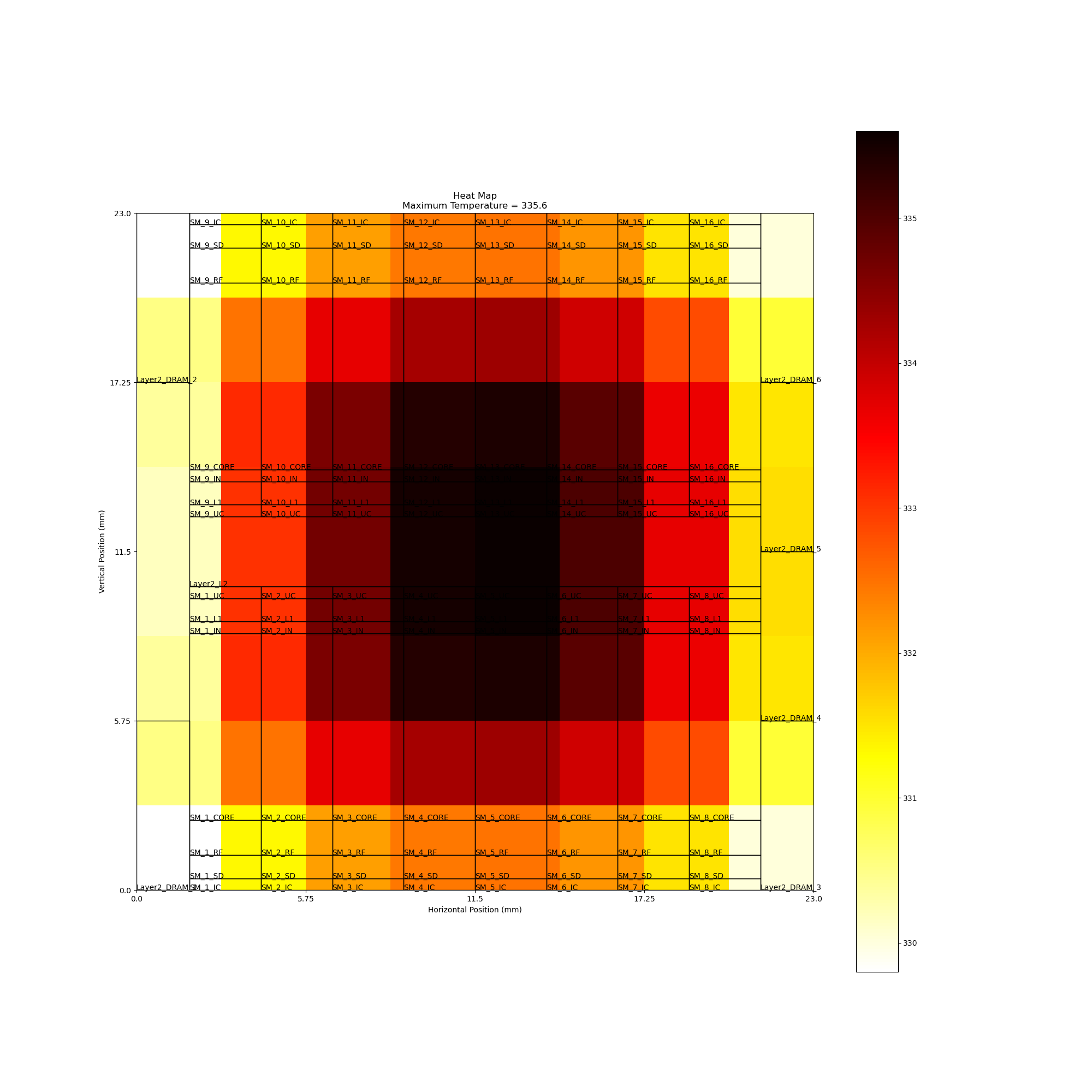}
    \caption{Cuda optimized Linearized Multiplication Kernel with reduced L2: 400x400}
    \label{fig:33}
\end{figure}

\begin{figure}[htpb]
    \centering
    \includegraphics[width=\linewidth]{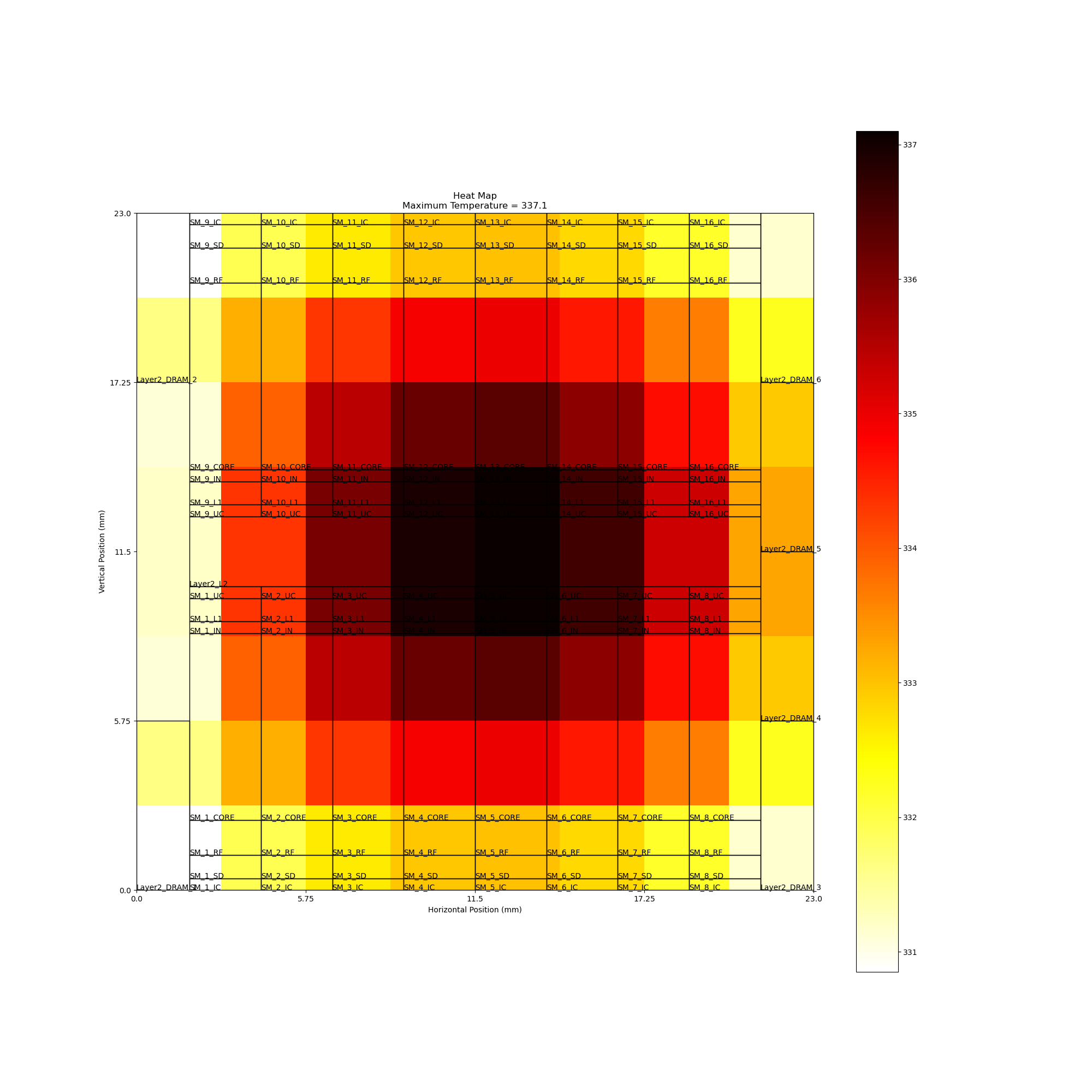}
    \caption{Cuda optimized Linearized Multiplication Kernel with reduced L2: 800x800}
    \label{fig:34}
\end{figure}

Figures \ref{fig:31} to \ref{fig:34} show the heat dissipation patterns for cuda optimized linearized multiplication kernel. The optimization involves using shared or scratchpad memory by a method called cache tiling. This leads to deterministic data accesses, where the tiles can now be conveniently stored in the private L2s of every SM. The experiments have been conducted with a reduced L2 cache size to observe pronounced patterns in DRAM accesses. Theoretically, since the data accesses are deterministic and efficient caching system can now be adapted, DRAM accesses have to minimized, thus leading to lesser DRAM heat dissipation. This has been observed clearly in the figures \ref{fig:31} to \ref{fig:34}. Comparing figure \ref{fig:24} and \ref{fig:34}, it is clearly seen that the DRAM access pattern with cuda optimization is as expected. Infact, the optimization is so powerful that the GPU is behaving as if the L2 cache size is the same as the original, i.e., the even with reduced cache size, DRAM accesses are minimal.  

\subsection{Needleman-Wunsch Kernel}
Needleman-Wunsch is a quadratic dynamic programming algorithm. Unlike matrix multiplication, the thread blocks are not independent. At any given time, for computing the score for $(i+1,j+1)$ index, we need data from rows $(i, j)$. This causal correlation between the thread blocks essentially reduces parallelization. As the tensor size increases, the DRAM gets significantly heated up, especially ones closest to the L2 cache. DRAM accesses is significantly more than that of linearized matrix multiplication kernel, for the tensor size of $800 \times 800$. This is because, same parts of L2 cache are being accessed each time by multiple thread blocks, leading to conflicts in L2 cache. Also, as the computation by different thread blocks is not independent, for memory consistency data needs to be constantly written to DRAM. 
\begin{figure}[htpb]
    \centering
    \includegraphics[width=\linewidth]{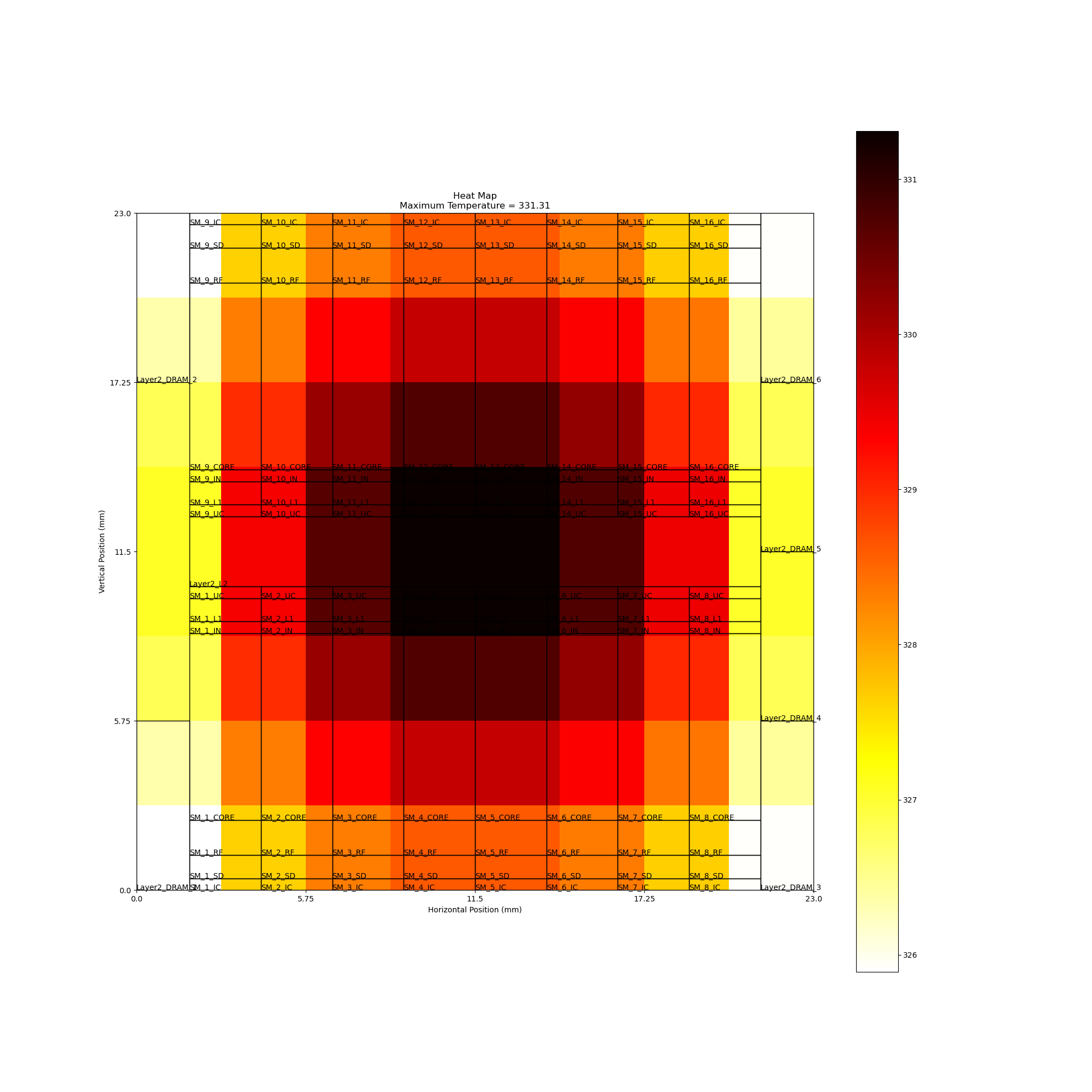}
    \caption{Needleman-Wunsch: 100x100}
    \label{fig:41}
\end{figure}
\begin{figure}[htpb]
    \centering
    \includegraphics[width=\linewidth]{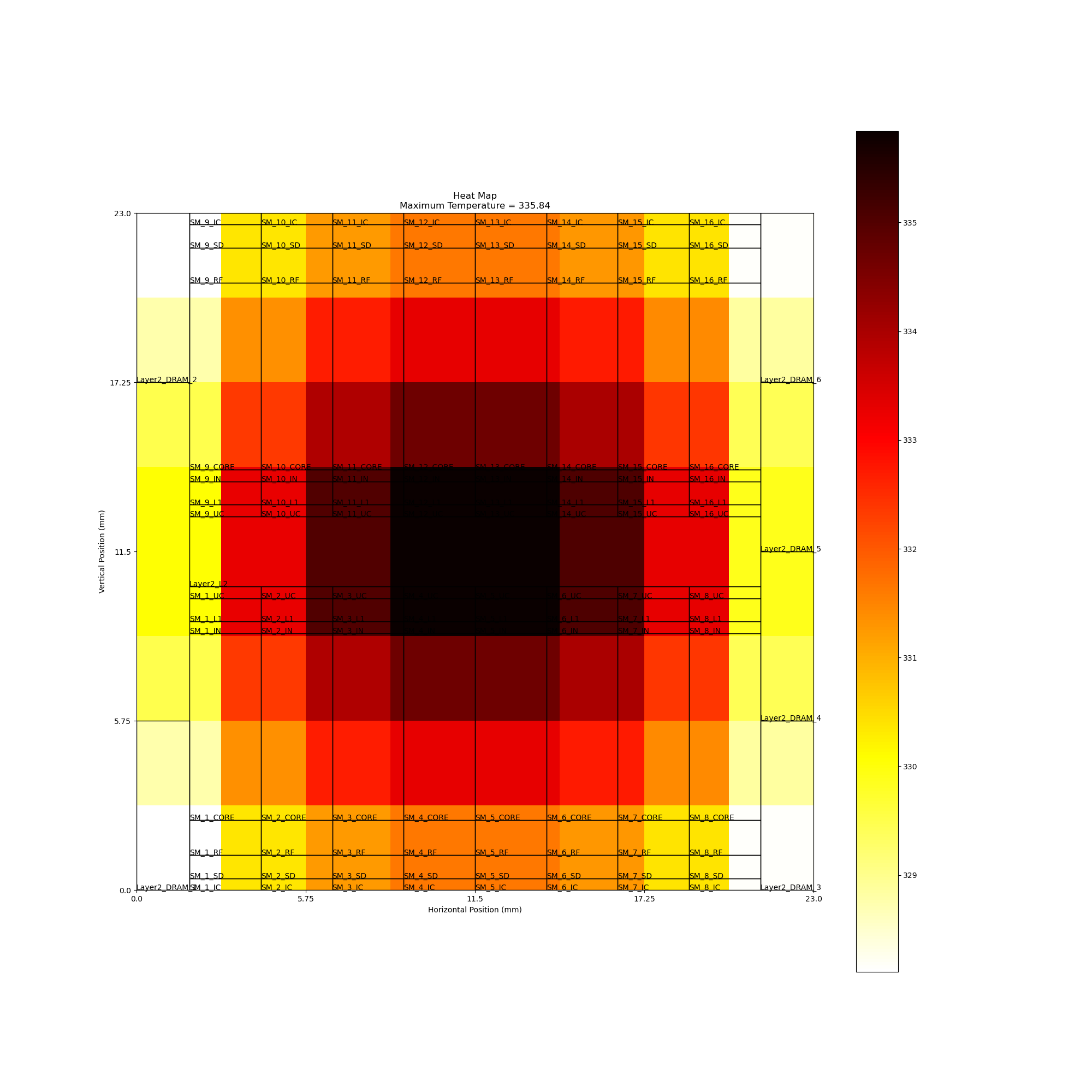}
    \caption{Needleman-Wunsch: 250x250}
    \label{fig:42}
\end{figure}
\begin{figure}[htpb]
    \centering
    \includegraphics[width=\linewidth]{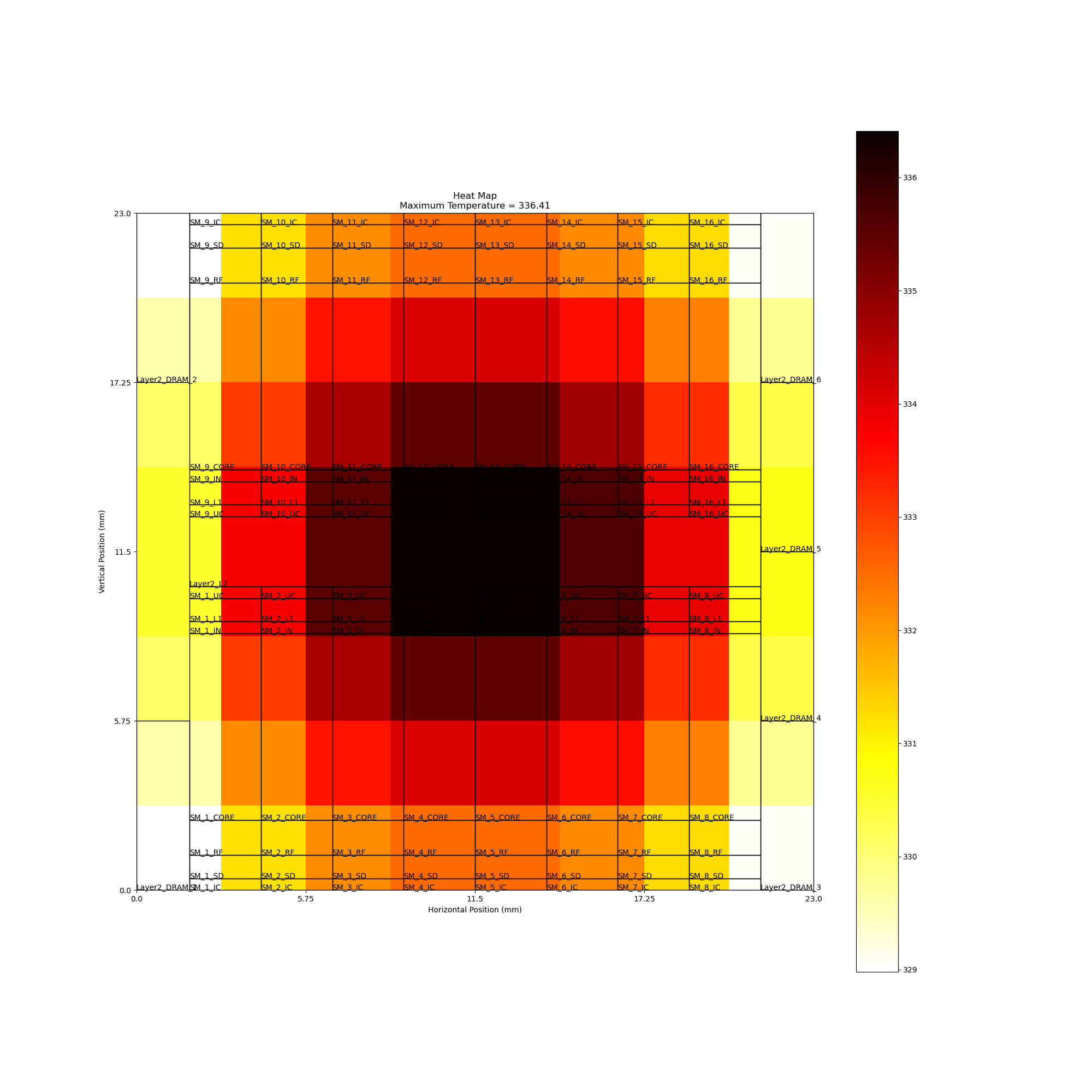}
    \caption{Needleman-Wunsch: 400x400}
    \label{fig:43}
\end{figure}
\begin{figure}[htpb]
    \centering
    \includegraphics[width=\linewidth]{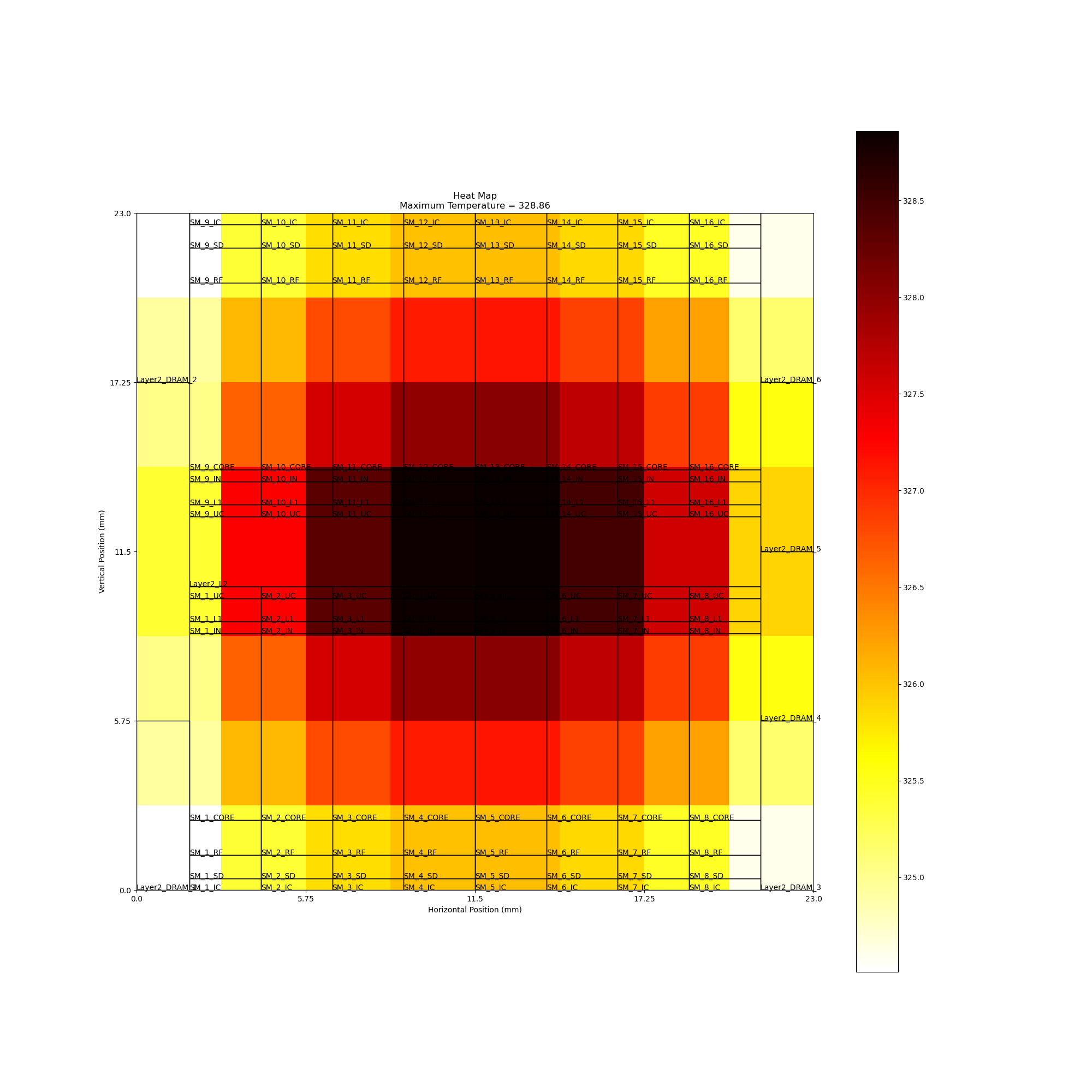}
    \caption{Needleman-Wunsch: 800x800}
    \label{fig:44}
\end{figure}
\begin{figure}[htpb]
    \centering
    \includegraphics[width=\linewidth]{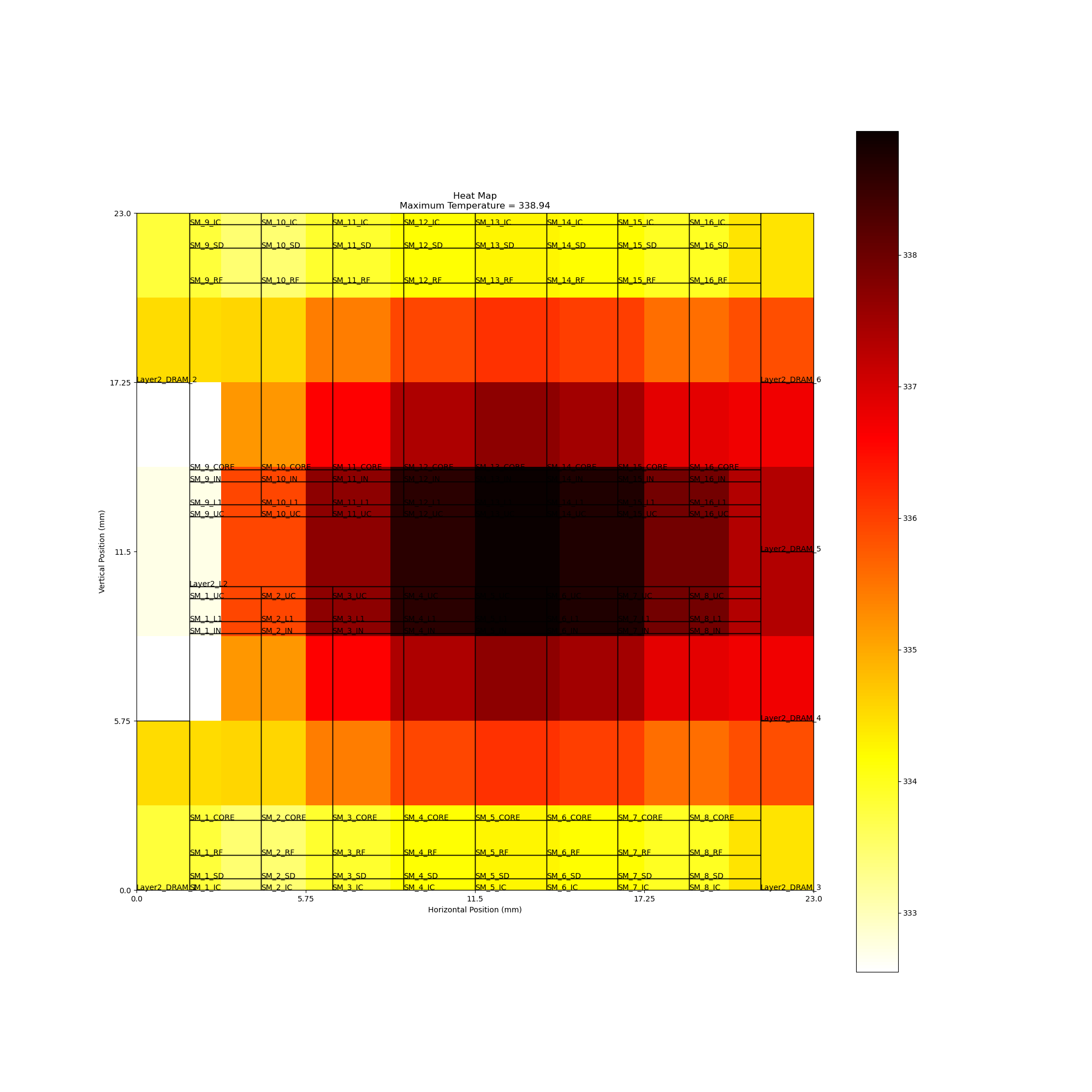}
    \caption{Needleman-Wunsch: 1200x1200}
    \label{fig:45}
\end{figure}

\subsection{Comparison between CUDA kernels}
Based on the heat map patterns, following are the conclusions we can draw regarding the CUDA kernels:
\begin{enumerate}
    \item Linearized matrix multiplication takes significantly more time to execute when compared to optimized linearized matrix multiplication. The computations are faster due to lesser data movement and deterministic data accesses in the L1 cache of each SM.  
    \item For the same tensor size, optimized linearized matrix multiplication significantly uses lesser DRAM accesses when compared to regular linearized matrix multiplication. This is because of the use of cache tiling or scratchpad memory.  
    \item Needleman-Wunsch kernel is memory intensive when compared to linearized matrix multiplication. In matrix multiplication, since the threads can work independently, and each subproblem is unrelated to the other, the L2 accesses will have lesser conflicts  when compared to Needlman-Wunsch, where the data in the left, top and diagonally top-left are necessary for an element's computation. This is why for a tensor-size of $800\times 800$, DRAM gets heated up more in Needleman-Wunsch when compared to matrix multiplication. 
\end{enumerate}

\section{Learning and Key Takeaways}
\begin{enumerate}
    \item This is the first time I am writing CUDA code. Initially, I was not aware of the benefits of linearization of matrices for GPU threads. Eventually, when I came accross linearization benefits, it was truly an eye-opener. I could run the kernel for higher tensor sizes with lesser time.
    \item CUDA optimization using the concept of scratchpad memory was very intriguing to me. I had a very good learning experience in applying what was learnt in class to a practical problem, and seeing it perform significantly better than non-optimized kernel in terms of the execution time and DRAM accesses. 
    \item Since I work in the field of machine learning, learning the intricacies and architecture of a GPU and being able to develop a floor plan was a great takeaway.
    \item I also learnt to use GPGPU-Sim, which was again a first time for me. I feel more confident in speaking about the practicals of computer architecture after the assignments and projects in this course.
\end{enumerate}
\section{Acknowledgements}
I would like to thank Prof. Vijay for the wonderful semester and intuitive classroom sessions. This is my final course of my PhD career, and it was definitely an amazing learning experience. 
{\small
\bibliographystyle{ieee_fullname}
\bibliography{egbib}
}

\end{document}